\documentclass[epj,nopacs]{svjour}

\usepackage[british]{babel}
\usepackage[utf8]{inputenc}
\usepackage{amsmath,amssymb,mathtools}
\usepackage{graphicx}
\DeclareGraphicsExtensions{.pdf,.png,.jpg}
\graphicspath{{./figures/}}
\usepackage[x11names,svgnames,table]{xcolor}
\usepackage{xspace}
\usepackage{booktabs,tabularx,multirow}
\usepackage{subcaption}
\usepackage[%
  citecolor=blue,%
  linkcolor=blue,%
  linktoc=page,%
  urlcolor=blue,%
  colorlinks=true%
]{hyperref}
\usepackage{cite}
\usepackage{siunitx}
\usepackage{listings}

\DeclareRobustCommand{\nnlojet}{\texorpdfstring{NNLO\protect\scalebox{0.8}{JET}}{NNLOJET}\xspace}
\DeclareRobustCommand{\pineappl}{\texorpdfstring{P\protect\scalebox{0.8}{INE}APPL}{PineAPPL}\xspace}
\DeclareRobustCommand{\ensuremathrm}[1]{\ensuremath{\mathrm{#1}}\xspace}

\DeclareRobustCommand{\LO}{\text{LO}\xspace}
\DeclareRobustCommand{\NLO}{\text{NLO}\xspace}
\DeclareRobustCommand{\NNLO}{\text{NNLO}\xspace}
\DeclareRobustCommand{\N}[1]{\ensuremath{\text{N}^{#1}}} 

\DeclareRobustCommand{\Pp}{\ensuremath{\mathrm{p}}\xspace}
\DeclareRobustCommand{\Pap}{\ensuremath{\bar{\mathrm{p}}}\xspace}
\DeclareRobustCommand{\Pg}{\ensuremath{\mathrm{g}}\xspace}
\DeclareRobustCommand{\Pq}{\ensuremath{q}\xspace}
\DeclareRobustCommand{\Paq}{\ensuremath{\bar{q}}\xspace}
\DeclareRobustCommand{\PZ}{\ensuremath{\mathrm{Z}}\xspace}

\DeclareRobustCommand{\PWm}{\ensuremath{\mathrm{W}^{-}}\xspace}

\DeclareRobustCommand{\fb}{\ensuremath{\mathrm{fb}}\xspace}
\DeclareRobustCommand{\rd}{\ensuremathrm{d}} 
\DeclareRobustCommand{\rT}{\ensuremathrm{T}} 
\DeclareRobustCommand{\mur}{\ensuremath{\mu_{\mathrm{R}}}\xspace}
\DeclareRobustCommand{\muf}{\ensuremath{\mu_{\mathrm{F}}}\xspace}
\DeclareRobustCommand{\as}{\ensuremath{\alpha_{\mathrm{s}}}\xspace}

\renewcommand\makeheadbox{%
  \hfill
  CERN-TH-2025-013
}

\begin{document}

\title{Fast interpolation grids for the Drell--Yan process}

\author{
  Juan Cruz--Martinez\inst{1}\and
  Alexander Huss\inst{1}\and
  Christopher Schwan\inst{2}
}

\institute{
  CERN, Theoretical Physics Department, CH-1211 Geneva 23, Switzerland \and
  Universität Würzburg, Institut für Theoretische Physik und Astrophysik, 97074 Würzburg, Germany
}

\date{Date: \today}

\abstract{
  Modern analyses of experimental data from hadron colliders rely on theory predictions at high orders in perturbation theory and a variety of input settings.
  Interpolation grids facilitate an almost instant re-evaluation of theory predictions for different input parton distributions functions (PDFs) or scale settings and are thus indispensable in the study of the parton content of the proton.
  While interpolation grids at next-to-next-to-leading order (\NNLO) exist for some key processes relevant for PDF determinations, a notable exception is the Drell--Yan process that constitutes the production of electroweak gauge bosons at hadron colliders and provides important constraints on the quark content of the proton.
  To address this gap, we report on a new interface between the parton-level Monte Carlo generator \nnlojet and the interpolation grid library \pineappl and demonstrate its use for the Drell--Yan process.
  Accompanying this note, we release Drell--Yan grids covering a wide range of measurements that commonly enter global determinations of PDFs.
  We use the grids to study accidental cancellation between partonic channels at \NNLO and inspect the validity of a $K$-factor approximation that was widely employed previously.
}

\maketitle

\section{Introduction}
\label{sec:intro}

\begin{sloppypar}
Over the past decades, the unprecedented precision achieved by the experiments at the Large Hadron Collider (LHC), as well as the projections of its high-luminosity upgrade (HL-LHC), has driven remarkable theoretical advances, pushing the frontier of perturbative Quantum Chromodynamics (pQCD) to next-to-next-to-leading order (\NNLO) accuracy and beyond.
Given the high computational resources consumed by such state-of-the-art calculations, efficient methods to re-evaluate predictions for different inputs are of critical importance.
Of particular interest are fast interpolation grids, which preserve the dependence of the partonic cross sections on the renormalisation and factorisation scales, the longitudinal momentum fractions carried by the partons within the colliding hadrons, and the strong coupling \as.
The generation and usage of interpolation grids are facilitated by libraries such as APPLgrid~\cite{Carli:2010rw}, fastNLO~\cite{Kluge:2006xs,Britzger:2012bs}, and \pineappl~\cite{Carrazza:2020gss,Schwan:2021txc}.
These grids enable fast evaluations of the cross section for different choices of parton distribution functions (PDFs) that would otherwise require a costly re-calculation, and are thus indispensable for the determination of PDFs.
\end{sloppypar}

Interpolation grids are available at \NNLO for various processes,
e.g., for jet production in deep-inelastic scattering (DIS)~\cite{Britzger:2019kkb,H1:2021xxi},
jet production in hadron collisions~\cite{Britzger:2022lbf},
top-quark pair production~\cite{Czakon:2017dip,Garzelli:2023rvx},
and inclusive DIS production~\cite{Bertone:2016lga,Candido:2024rkr}.
A notable absence in this list is the Drell--Yan (DY) process, which constitutes the production of electroweak gauge bosons at hadron colliders.
While grids were made available for inclusive fixed-target DY~\cite{Barontini:2023vmr}, no differential predictions are available despite it being a process of high importance for PDF determinations, accounting for roughly 20\% of the total dataset in recent global fits~\cite{Hou:2019efy,Bailey:2020ooq,NNPDF:2021njg}.

Moreover, the Drell--Yan process~\cite{Drell:1970wh} is a high-precision probe at hadron colliders and it has been known fully differentially at \NNLO for a long time~\cite{Anastasiou:2003ds,Melnikov:2006kv,Catani:2009sm,Catani:2010en}, and most recently also computed at \N3\LO both at the inclusive~\cite{Duhr:2020seh,Duhr:2020sdp,Duhr:2021vwj,Baglio:2022wzu} and differential level~\cite{Chen:2021vtu,Chen:2022cgv,Chen:2022lwc,Neumann:2022lft,Campbell:2023lcy}.
Despite the availability of predictions, no differential grids are available at \NNLO and so current PDF determinations commonly rely on an approximation basted on \NLO grids supplemented with \NNLO $K$-factors obtained for a specific choice of PDF.
In this note we interface the parton-level event generator \nnlojet~\cite{NNLOJET:2025xyz} that is based on the antenna subtraction formalism~\cite{Daleo:2006xa,Boughezal:2010mc,Gehrmann:2011wi,Gehrmann-DeRidder:2012too,Gehrmann-DeRidder:2005btv} with the \pineappl~\cite{Carrazza:2020gss,Schwan:2021txc} grid-interpolation library.
This enables the production of interpolation grids for all processes included in the \nnlojet code in a variety of scenarios.
We demonstrate the use of this new interface using the DY process as an example.
Together with this paper we make available a wide range of grids that correspond to the entire set of DY data that enters the NNPDF4.0~\cite{NNPDF:2021njg} determination of PDFs.
The grids are available to download at:%
\footnote{
  Grids are provided in the \pineappl format; conversion \emph{from} and \emph{to} (\lstinline{pineappl [import|export] --help}) this format facilitate its use as an universal converter.
}
\begin{center}
  \url{https://ploughshare.web.cern.ch}
\end{center}

This note is structured as follows: in Section~\ref{sec:grids} we describe the grids that we make available and provide a summary of the metadata that is included.
We demonstrate the accuracy of the interpolation and provide some examples of studies that can be readily performed with the \pineappl command-line interface (CLI).
We end in Section~\ref{sec:pheno} with new insights into the known feature of accidental cancellations at NNLO and assess the impact of the widely used $K$-factor approximation in the context of global PDF analyses.

\section{Interpolation grids}
\label{sec:grids}

\begin{table*}



\begin{tabularx}{\textwidth}{Xccc}
  \toprule
  Dataset
  & Ref.
  & $N_{\rm dat}$
  & Cuts
  \\
  \midrule
  CDF $Z$ differential
  & \cite{CDF:2010vek}
  & 29
  & $0.0\le y_{\ell\ell}\le 2.9$, $66\le m_{\ell\ell}\le 116$
  \\
  D0 $Z$ differential
  & \cite{D0:2007djv}
  & 28
  & $0.0\le y_{\ell\ell}\le 2.8$, $66\le m_{\ell\ell}\le 116$
  \\
  D0 $W$ electron asymmetry
  & \cite{D0:2013xqc}
  & 13
  & $0.0\le y_{e}\le 2.9$
  \\
  D0 $W$ muon asymmetry
  & \cite{D0:2014kma}
  & 10
  & $0.0\le y_{\mu}\le 1.9$
  \\  
  \midrule
  ATLAS low-mass DY 7 TeV
  & \cite{ATLAS:2014ape}
  & 6
  & $|\eta_\ell|\leq 2.1$, $14\le m_{\ell\ell}\le 56$
  \\
  ATLAS high-mass DY 7 TeV
  & \cite{ATLAS:2013xny}
  & 13
  & $|\eta_\ell|\leq 2.1$, $116\le m_{\ell\ell}\le 1500$
  \\
  ATLAS $W,Z$ 7 TeV ($\mathcal{L}=35$~pb$^{-1}$)
  & \cite{ATLAS:2011qdp}
  & 30
  & $|\eta_\ell,y_Z|\leq 3.2$
  \\
  ATLAS $W,Z$ 7 TeV ($\mathcal{L}=4.6$~fb$^{-1}$)
  & \cite{ATLAS:2016nqi}
  & 61
  & $|\eta_\ell,y_Z|\leq 2.5,3.6$
  \\
  ATLAS $W$ 8 TeV
  & \cite{ATLAS:2019fgb}
  & 22
  & $|\eta_\ell|<2.4$
  \\
  ATLAS low-mass DY 2D 8 TeV
  & \cite{ATLAS:2017rue}
  & 84
  & $|y_{\ell\ell}|<2.4$, $46\le m_{\ell\ell}\le 200$
  \\
  ATLAS high-mass DY 2D 8 TeV
  & \cite{ATLAS:2016gic}
  & 48
  & $|y_{\ell\ell}|<2.4$, $116\le m_{\ell\ell}\le 1500$
  \\
  CMS $W$ electron asymmetry 7 TeV
  & \cite{CMS:2012ivw}
  & 11
  & $|\eta_e|\leq 2.4$
  \\
  CMS $W$ muon asymmetry 7 TeV
  & \cite{CMS:2013pzl}
  & 11
  & $|\eta_\mu|\leq 2.4$
  \\
  CMS DY 2D 7 TeV
  & \cite{CMS:2013zfg}
  & 132
  & $|\eta_{\ell\ell}|\leq 2.2$, $20.0\leq m_{\ell\ell}\leq 200$
  \\
  CMS $W$ rapidity 8 TeV
  & \cite{CMS:2016qqr}
  & 22
  & $|\eta_\ell|\leq 2.3$
  \\
  LHCb $Z\to ee$ 7 TeV 
  & \cite{LHCb:2012gii}
  & 9
  & $2.0\leq \eta_\ell\leq 4.5$
  \\
  LHCb $W,Z \to \mu$ 7 TeV
  & \cite{LHCb:2015okr}
  & 33
  & $2.0\leq \eta_\ell\leq 4.5$
  \\
  LHCb $Z\to ee$ 8 TeV
  & \cite{LHCb:2015kwa}
  & 17
  & $2.00<|\eta_e|<4.25$
  \\
  LHCb $W,Z\to \mu$ 8 TeV
  & \cite{LHCb:2015mad}
  & 34
  & $2.00<|\eta_\mu|<4.25$
  \\
  LHCb $W \to e$ 8 TeV
  & \cite{LHCb:2016zpq}
  & 8
  & $2.00<|\eta_e|<4.25$
  \\
  ATLAS $\sigma_{W,Z}^{\rm tot}$ 13 TeV
  & \cite{ATLAS:2016fij}
  & 3
  & ---
  \\
  LHCb $Z\to ee$ 13 TeV
  & \cite{LHCb:2016fbk}
  & 17
  & $2.00<|y_Z|<4.25$
  \\
  LHCb $Z\to \mu\mu$ 13 TeV
  & \cite{LHCb:2016fbk}
  & 18
  & $2.00<|y_Z|<4.50$
  \\
  \bottomrule
\end{tabularx}

  \caption{Collider Drell--Yan datasets considered in the NNPDF4.0 global PDF determination and selected for this release of interpolation grids.
  Table of kinematic cuts from Ref.~\cite{NNPDF:2021njg}.}
  \label{tab:dataset}
\end{table*}

The general idea of interpolation grids follows from the property of QCD factorisation, where hadronic cross sections can be written in terms of a convolution of the partonic cross section with the PDFs.
Approximating the PDFs through a decomposition into a set of eigenfunctions, a grid can be produced that corresponds to the convolution of said eigenfunctions with the partonic cross section.
With this at hand, the evaluation of a hadronic cross section reduces to performing a sum over the grid entries weighted by the PDFs at the given nodes, thus, substantially speeding up the evaluation.
Further details on the grid-interpolation techniques as implemented in the \pineappl library can be found in Ref.~\cite{Carrazza:2020gss}.

Accompanying this manuscript, we release a large set of interpolation grids for the Drell--Yan process corresponding to the measurements summarised in Table~\ref{tab:dataset} that were produced using the implementation of this process within the \nnlojet framework~\cite{Gehrmann-DeRidder:2023urf}.
All predictions are produced for the central scale choice \(\mur=\muf=E_{\rT,V}\), where the transverse energy \(E_{\rT,V}\) is defined in terms of the invariant mass of the intermediate electroweak gauge boson and its transverse momentum, \(E_{\rT,V} = \sqrt{m_{V}^2 + p_{\rT,V}^2}\).
This choice largely coincides with the invariant mass \(m_{V}\) in inclusive quantities, however, it accounts for the impact of hard QCD emissions in phase-space regions that are sensitive to it.

Lastly, all grids evaluate to absolute predictions in \fb units with appropriate bin-width normalisations following the HEPData entries provided with the corresponding measurements.
No composition of observables is performed in order to maintain the full flexibility and granularity that these grids offer.
In particular, this means that in order to obtain normalised distributions, such as $(1/\sigma)\, \rd \sigma / \rd \mathcal{O}$, this has to be performed by the user by appropriately summing the cross section of the individual bins.

\subsection{Metadata}
\label{sec:metadata}

The interpolation grids made available alongside this publication contain the following metadata that are accessible through the \pineappl CLI:%
\footnote{\lstinline{pineappl read <grid> --get <key>}}

\begin{description}
  \item[\texttt{nnlojet\_runcard} ---] A sample \nnlojet runcard used to produce the grid.
  \item[\texttt{nnlojet\_version} ---] Version of \nnlojet\ used.
  \item[\texttt{pineappl\_gitversion} ---] Version of \pineappl used.
  \item[\texttt{result} ---] Reference numbers as reported by \nnlojet broken down into different perturbative orders together with associated Monte Carlo integration errors.
  \item[\texttt{results\_pdf} ---] The PDF set used to generate the results. This information is necessary in order to perform closure tests shown in Section~\ref{sec:gridclosure}.
  \item[\texttt{hepdata} ---] HEPData entry of the dataset for which the grids were produced, c.f.\ Table~\ref{tab:dataset}.
\end{description}

\subsection{Closure tests}
\label{sec:gridclosure}

\begin{figure*}
  \centering
  \begin{subfigure}{0.48\linewidth}
    \includegraphics[width=\linewidth]{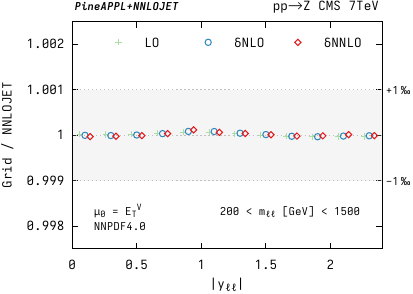}
    \caption{closure test for the setup of Ref.~\cite{CMS:2013zfg}}
    \label{fig:gclosure_one}
  \end{subfigure}
  \hfill
  \begin{subfigure}{0.48\linewidth}
    \includegraphics[width=\linewidth]{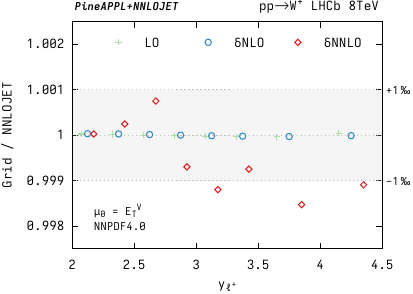}
    \caption{closure test for the setup of Ref.~\cite{LHCb:2015mad}}
    \label{fig:gclosure_two}
  \end{subfigure}
  \caption{
    Grid closure between the interpolation and the exact \nnlojet reference numbers for the datasets of Refs.~\cite{CMS:2013zfg}~(\subref{fig:gclosure_one}) and \cite{LHCb:2015mad}~(\subref{fig:gclosure_two}) at LO and the NLO and NNLO coefficients.
  }
  \label{fig:gclosure}
\end{figure*}

One of the main advantages of interpolation grids is the relatively small storage requirement associated with them.
This is in contrast to approaches based on storing separate collision events, such as Ntuples~\cite{Bern:2013zja,Maitre:2020blv} or HighTEA~\cite{Czakon:2023hls}, which often have a significantly larger storage footprint.
The smaller size is achieved by both fixing the setup (event selection and binning of histograms) and thus giving up on some flexibility, as well as the approximation through interpolation.
The latter demands that any systematic uncertainty introduced by the interpolation to be well below other uncertainties in the calculation.
In the case of the Drell--Yan process, where \NNLO corrections are known to be very small and plagued by large numerical cancellations, assessing the quality of the grids is particularly important.

In Fig.~\ref{fig:gclosure} we present closure tests that compare predictions obtained from evaluating interpolation grids against exact reference numbers from the calculation used to generate them.
Any observed difference is thus solely due to interpolation errors.
To this end, we decompose the \NNLO predictions into separate contributions
\begin{align*}
  \rd\sigma^{\NNLO} &=
  \rd\sigma^{\LO} + \rd\sigma^{\delta\NLO} + \rd\sigma^{\delta\NNLO}
  \; ,
\end{align*}
and perform the closure test on each of the perturbative \emph{coefficients}.
Figure~\ref{fig:gclosure_one} highlights that the interpolation errors are typically well below the per-mille level in the bulk of the phase space.
However, deviations can become as large as few per-mille in extreme cases such as the forward region of the LHCb experiment that probes a much larger range of the momentum fractions as seen in Fig.~\ref{fig:gclosure_two}.
It should be emphasized that \(\rd\sigma^{\delta\NNLO}\) typically amounts to a few percent of the full \NNLO prediction and an interpolation error of 1\textperthousand\ on the coefficient translates to a 0.01\textperthousand\ level of exactitude on the final results.
In contrast, the residual Monte Carlo errors on the predictions are at the few \textperthousand\ level, which in turn are already negligible with respect to the experimental uncertainties of the measurement.
For any phenomenological application, the impact of interpolation errors in the provided grids is thus completely negligible.

\subsection{Uncertainties}
\label{sec:uncertainties}

A common use-case for grid files is the study of various sources of uncertainties that require re-evaluating predictions for different input parameters.
The \pineappl CLI provides efficient tools for analysing grids in various scenarios.
Below, we demonstrate some representative analyses using the built-in utilities.

\paragraph{Scale variations}
are the most widely used approach to estimate theory uncertainties from missing higher orders in the perturbative expansion.
The \pineappl grids allow to vary the renormalisation and factorisation scales, \mur and \muf respectively, through any linear transformation of the original scale used for the grid generation.
This offers the possibility to explore the \((\mur,\muf)\) space with high granularity.

\begin{figure}
  \centering
  \includegraphics[width=0.95\linewidth]{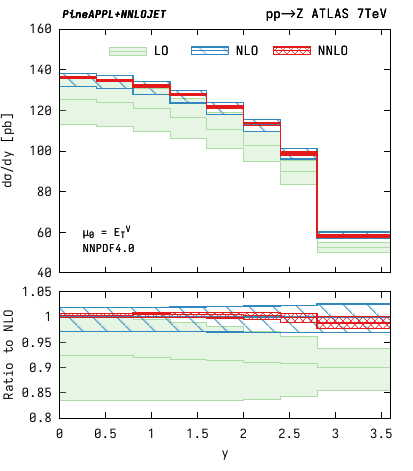}
  \caption{9-point scale variations for the Z data from Ref.~\cite{ATLAS:2011qdp}, evaluated using the PDF set (\texttt{NNPDF40\_nnlo\_as\_01180}) at all orders.}
  \label{fig:pine_scale}
\end{figure}

The conventional variation by factors of \((\frac{1}{2}, 2)\) is directly provided within the CLI through the command%
\footnote{Instead of the 9-point variation, also the 3- and 7-point variations are available, which can be specified through the \lstinline{--scale-abs} option.}
\begin{lstlisting}
  pineappl uncert <grid> <pdf>
      --scale-abs=9 --orders <orders>
\end{lstlisting}
where \lstinline{<orders>} specifies the perturbative orders to be considered, e.g.\ \lstinline{--orders a2,as1a2} for NLO in the DY process.
In Fig.~\ref{fig:pine_scale} we show as an example the 9-point scale variation uncertainties for the Z production calculation for the measurement of Ref.~\cite{ATLAS:2011qdp}.

\paragraph{PDF uncertainties}
are derived from a collection of predictions obtained via a convolution with PDFs of an associated error set.
Interpolation grids provide detailed information across several parameters, not only the renormalisation and factorisation scales, to facilitate the efficient a posteriori re-evaluation of the predictions for arbitrary PDF sets and thus the study of PDF uncertainties as well as their use in PDF fits.
To this end, interpolation grids additionally retain the information on the longitudinal momentum fractions of the colliding partons, \(x_1\) and \(x_2\), separately for each independent combination of partonic channels of the corresponding process.

\begin{figure}
  \centering
  \includegraphics[width=0.95\linewidth]{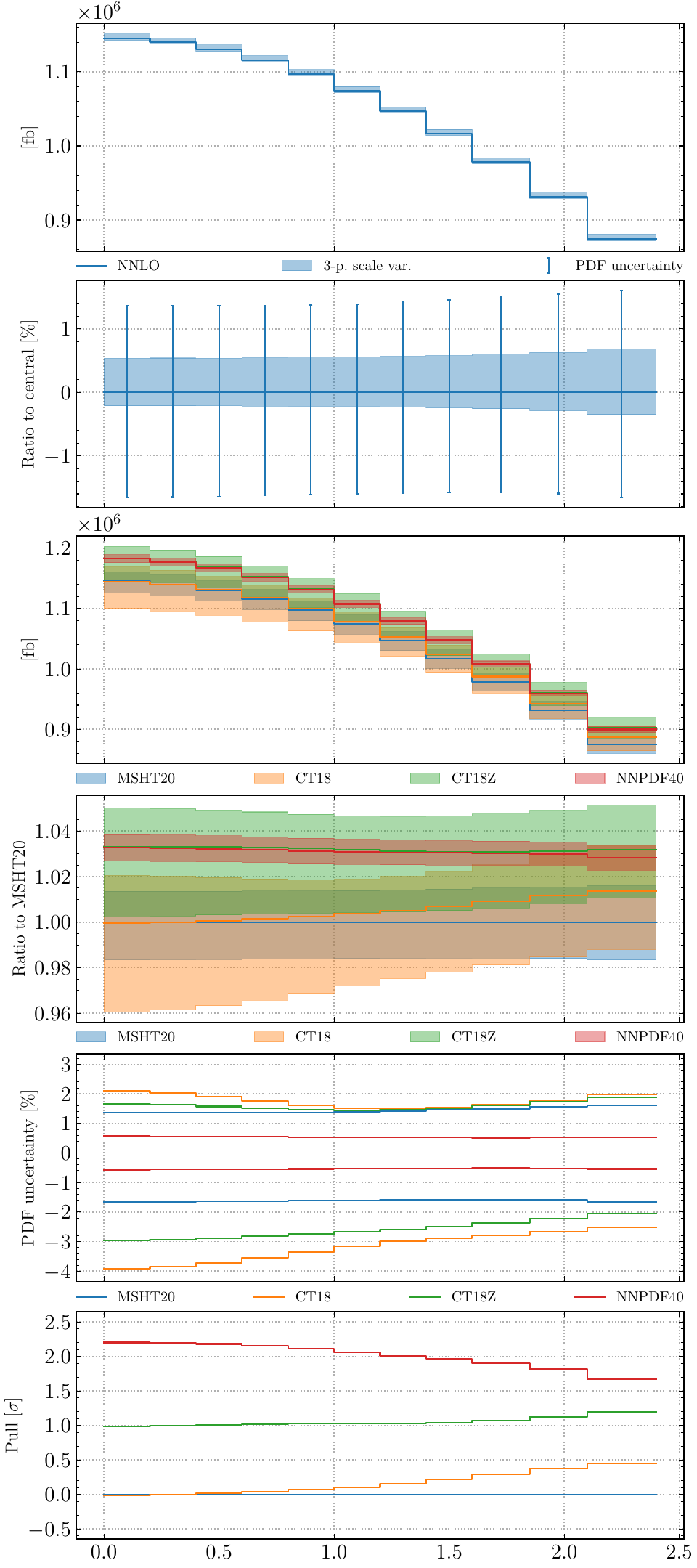}
  \caption{Scale and PDF uncertainties for the \PWm data of Ref.~\cite{CMS:2016qqr}. Using interpolation grids we can obtain several different analyses with a single command of \texttt{PineAPPL} in a matter of seconds.}
  \label{fig:pine_pdf}
\end{figure}

In Fig.~\ref{fig:pine_pdf}, we contrast the PDF uncertainties of different sets at NNLO from the MHST~\cite{Bailey:2020ooq}, CTEQ~\cite{Hou:2019efy} and NNPDF~\cite{NNPDF:2021njg} fitting groups.
The figure is obtained with the CLI command
\begin{lstlisting}
  pineappl plot -s 3 <grid> <pdf1> <pdf2>...
\end{lstlisting}
which is used together with the option \lstinline{-s 3} to specify the 3-point scale variation in this case.
The output of the command is a Python script to generate a figure as shown in Fig.~\ref{fig:pine_pdf} and is composed of several panels:
The first two panels show separately the scale- and PDF-uncertainties of the predictions using \lstinline{<pdf1>} (in this case MSHT20) for the absolute prediction and the relative uncertainties.
This is followed by three panels that show the PDF uncertainties for all PDF sets that were specified, which are provided as absolute predictions and ratio plots with respect to the central \lstinline{<pdf1>}.
The final two panels directly contrast the PDF uncertainties of the different sets together with the pull in units of $\sigma$ for each PDF, using \lstinline{<pdf1>} as reference.

By utilising interpolation grids, analyses can be made with multiple PDF set at zero added computational cost.
The importance of these extra studies is clear e.g., in the 4th and 6th panels of Fig.~\ref{fig:pine_pdf}.
While the pull between sets can be as large as \(2\sigma\), and thus the choice of PDF might have phenomenological implications, this difference is washed out once several sets are included in the analysis.

\section{Drell--Yan phenomenology}
\label{sec:pheno}

\subsection{Accidental cancellations}
\label{sec:cancellation}

\begin{figure*}
  \centering
  \begin{subfigure}{0.48\linewidth}
    \includegraphics[width=\linewidth]{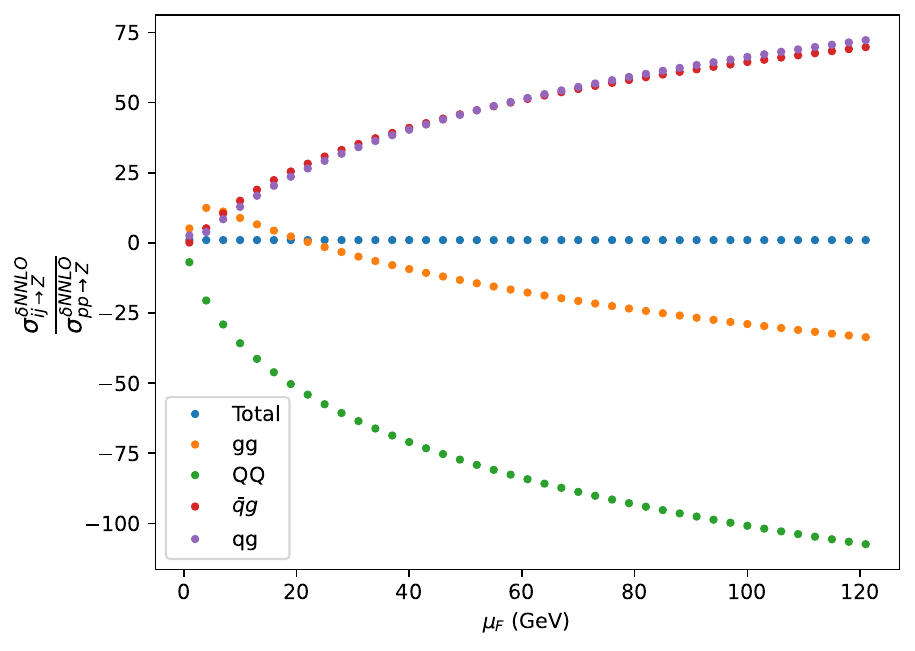}
    \caption{flavour basis}
    \label{fig:cancellation_flav}
  \end{subfigure}
  \hfill
  \begin{subfigure}{0.48\linewidth}
    \includegraphics[width=\linewidth]{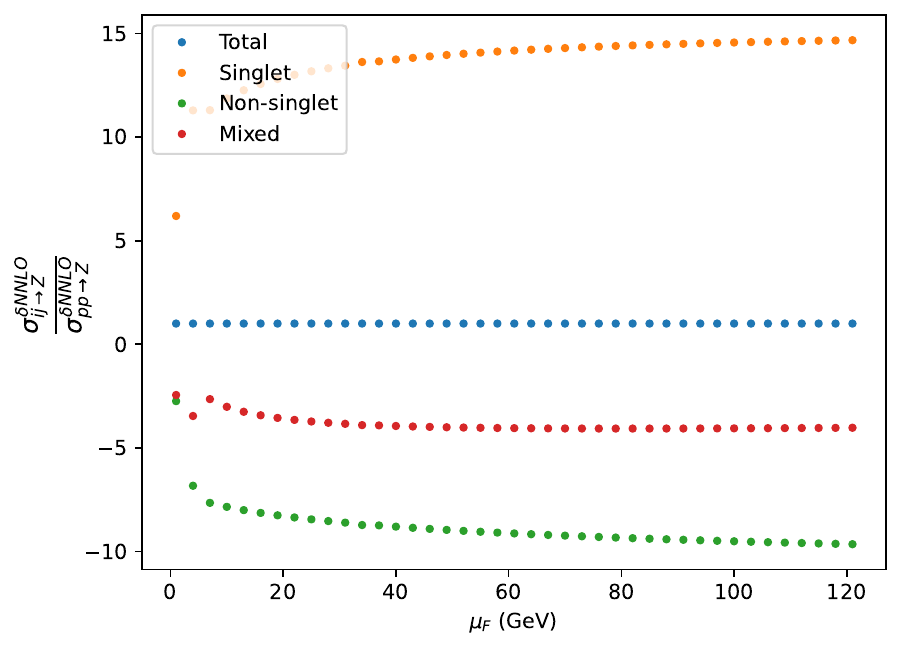}
    \caption{evolution basis}
  \label{fig:cancellation_evol}
  \end{subfigure}
  \caption{
    Channel decomposition of the NNLO contribution \(\sigma^{\delta\NNLO}\) to the total cross section for \PZ production as a function of the scale, normalised to the full NNLO contribution.
    Decomposition is shown both for the flavour basis~(\subref{fig:cancellation_flav}) and the evolution basis~(\subref{fig:cancellation_evol}).
    The ``$QQ$'' channel includes all quark and anti-quark initiated contributions.}
  \label{fig:cancellation}
\end{figure*}

One peculiar feature of the DY process that has been the subject of various investigations are the large accidental cancellation between the \(\Pq\Paq\) and \((\Pq\Pg+\Paq\Pg)\) partonic channels.
This cancellation is especially severe at \NNLO, in part causing the scale uncertainties to be underestimated and for the \N3\LO predictions to often lie outside of the \NNLO uncertainty estimates~\cite{Duhr:2020seh}.
This striking trait of non-overlapping scale-uncertainty bands is particularly pronounced at higher hadron-collider energies and was found to be largely independent of whether \(\Pp\Pp\) or \(\Pp\Pap\) collisions are considered~\cite{Baglio:2022wzu}.
These observations hint towards cancellations that are likely driven by correlations between the gluon and sea-quark distributions.

\begin{sloppypar}
Interpolation grids offer the unique opportunity to study such hypotheses in more detail by providing the possibility to evaluate the predictions at different orders and broken apart into individual partonic channels (using different bases), combined with the flexibly to evolve the prediction to any value for the scales.
Practically, this is achieved by evolving the grids with the DGLAP evolution library EKO~\cite{candido_2022_6340153,Candido:2022tld} to different values in \muf, a feature provided by the \pineappl CLI:%
\footnote{
  Note that it might be necessary to set a high value for the accuracy if the \lstinline{<pdf_to_check>} used for the check has not been evolved with the exact same settings as the kernel operator.
}
\begin{lstlisting}
  pineappl evolve <grid>
      <evolution_kernel_operator.tar>
      <output_evolved_grid>
      <pdf_to_check> --orders a2as2
\end{lstlisting}
where the evolution kernel operator is generated with the \texttt{pineko} program~\cite{Barontini:2023vmr} and the option \lstinline{--orders a2as2} ensures that only the NNLO QCD contribution is being considered.
\end{sloppypar}

Figure~\ref{fig:cancellation_flav} shows the NNLO contribution \(\sigma^{\delta\NNLO}\) separated into partonic channels as a function of the factorisation scale \muf.
The results correspond to the total cross section obtained with the setup of the measurement in Ref.~\cite{ATLAS:2016fij}.
The figure clearly highlights the large cancellation between the \(\Pq\Paq\) and \((\Pq\Pg+\Paq\Pg)\) channels, with a compensation of almost two orders of magnitude around \(\muf\sim M_\PZ\).
We further observe a strong dependence on the factorisation scale that enhances the cancellation for larger \muf values, which hints at an underlying correlation in the DGLAP evolution that drives this feature.

In order to gain further insights into this correlation and its impact, we transform to the so-called evolution basis%
\footnote{
  The exact definition of the evolution basis is provided in the \href{https://eko.readthedocs.io/en/latest/theory/FlavorSpace.html\#qcd-evolution-basis}{EKO documentation}.
}
by rotating the evolution operator with EKO, which serves the purpose of decoupling the evolution of the different independent components.
In particular, the gluon evolves together with the combination of quark distributions \(\Sigma = \sum_{i=1}^{6} (\Pq_{i}+\Paq_{i})\) as the so-called singlet contribution.
The non-singlet part includes contributions that in EKO follow the common notation of $V_i$ and $T_i$, while ``mixed'' refers to the contribution where both a singlet and non-singlet part enter the convolution.
The dependence on \muf for a decomposition in this basis is shown in Fig.~\ref{fig:cancellation_evol}.
Indeed, in this basis two points are immediately apparent:
first, the cancellation is much less pronounced even at high values of \muf by approximately an order of magnitude and; second, the dependence on the factorisation scale is rendered almost flat.
This demonstrates that much of the cancellations occurs \emph{within} the singlet sector and that the large cancellations observed in the flavour basis are largely an artefact of the choice of basis and the correlations introduced by the DGLAP evolution.

\subsection{Stability of \texorpdfstring{\(K\)}{K}-factors}
\label{sub:K-fac}

\begin{sloppypar}
The accidental cancellation between partonic channels highlighted in the previous section could, in principle, be either amplified or diminished by the choice of PDF, as they control the relative size of each partonic channel.
In turn one might expect the NNLO contribution to the cross section to be significantly impacted by the choice of PDF.
\end{sloppypar}

\begin{figure}
  \centering
  \begin{subfigure}{0.95\linewidth}
    \includegraphics[width=0.95\linewidth]{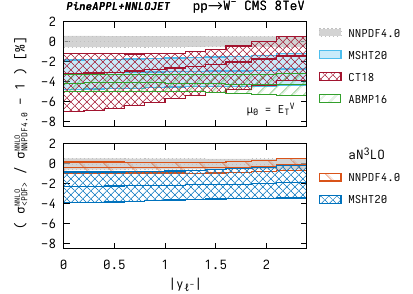}
    \caption{cross sections}
    \label{fig:nnlo_pdfs}
  \end{subfigure}
  \\\bigskip
  \begin{subfigure}{0.95\linewidth}
    \includegraphics[width=0.95\linewidth]{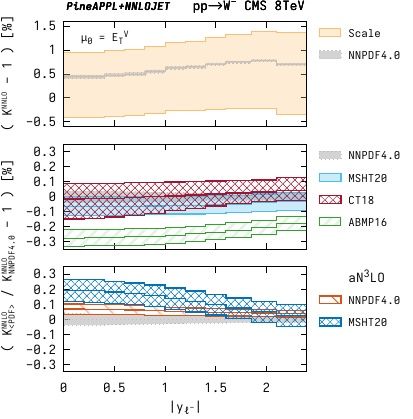}
    \caption{\(K\)-factors}
    \label{fig:nnlo_k}
  \end{subfigure}
  \caption{Comparison of different PDF sets for NNLO cross sections~(\subref{fig:nnlo_pdfs}) and \(K\)-factors~(\subref{fig:nnlo_k}) in units of \% with respect to the NNPDF4.0 set.}
  \label{fig:kfactors}
\end{figure}

Figure~\ref{fig:nnlo_pdfs} provides a detailed comparison of \NNLO predictions for a variety of PDF sets, where the bands correspond to the respective PDF uncertainties.
In the bottom panel, predictions are also shown for approximate \N3\LO sets in order to assess the impact from the (approximate) higher-order DGLAP evolution implemented in these sets.
The latter is particularly important as the previous section exposed the correlation induced by the evolution to be a main driver of the cancellations observed at \NNLO.
The spread between the different PDFs is at the few percent level and largely covered by the respective uncertainties.
The largest deviations are seen between the NNPDF and ABM sets of about 4\%, which can be attributed to the different datasets included in the fits and the different methodologies employed rather than the details of this calculation.
The impact of the approximate \N3\LO evolution is found to be small and well within the PDF uncertainties.

Without the availability of \NNLO grids, approximations were constructed based on \NLO grids supplemented by \NNLO \(K\)-factors that were obtained for a specific choice of PDF,
\begin{align*}
  \sigma^{\NNLO}
  &\approx \sigma^{\NLO} \times K^{\NNLO}_\mathrm{PDF}
  ,&
  K^{\NNLO}_\mathrm{PDF}
  &\equiv \frac{\sigma^{\NNLO}_\mathrm{PDF}}{\sigma^{\NLO}_\mathrm{PDF}}
  \; .
\end{align*}
For this approximation to be valid, the relevant property to inspect is not the total cross section but the \(K\)-factors and how stable they are with respect to the choice of the underlying PDF set used.
In Fig.~\ref{fig:nnlo_k} we show the \(K\)-factors for the same PDF sets as in Fig.~\ref{fig:nnlo_pdfs} including in the top panel a comparison of the (correlated) PDF uncertainties against the size of the \NNLO scale variation.
With the \(K\)-factors typically at the percent level, we observe that variation of the \(K\)-factors under the change of the PDF sets is very stable and only at the level of a few per-mille.
The final impact on the full \NNLO cross section \(\sigma^\NNLO\) from the \(K\)-factor approximation is thus estimated to be at the order of 0.01\textperthousand\ and thus negligible.

\subsection{\texorpdfstring{\(K\)}{K}-factor approximation and PDF fits}
\label{sec:pdf_fits}

Given the current claimed level of accuracy of PDF global analyses (\N3\LO, albeit approximated)~\cite{McGowan:2022nag,NNPDF:2024nan} and the uncertainty achieved in regions well constrained by data (close to 1\%)~\cite{NNPDF:2021njg} it is critical to study and remove any possible sources of bias in order to ensure the accuracy and robustness of the PDFs and its associated uncertainties.
While the previous section supports the robustness of the \(K\)-factor approximation for the DY process at the level of the full cross section, it is important to also verify the impact of this approximation on global PDF fits.
This is because PDF fits are sensitive to corrections for the \emph{individual} channels that are not correctly captured by a global \(K\)-factor nor are partonic channels that only open up at \NNLO (such as \(\Pg\Pg\) and \(\Pq\Pq'\)).

Moreover, PDF analyses are among the main use-cases of interpolation grids:
Determining PDFs relies heavily on comparing data and predictions for many observables and varying inputs.
This process involves evaluating hundreds of differential cross-sections at the PDF
fitting scale for each step in the optimisation procedure.
A global analysis of PDFs requires computing predictions for approximately 5000 datapoints to NNLO~\cite{NNPDF:2021njg}.
Given the complexity and the computational cost, handling this vast amount of data without the speed-up that the grids provide is practically infeasible.

\begin{sloppypar}
Since grids include detailed information about scales and orders, they can be convolved with evolution operators, to generate optimized Fast-Kernel tables.
These tables optimize away information about the orders and scales such that they can be directly convolved with PDFs at the fitting scale~\cite{Bertone:2016lga}.
This makes them particularly well suited for PDF determination. We have used the tools outlined in Ref.~\cite{Barontini:2023vmr} to prepare Fast-Kernel tables for all grids provided together with this paper.
\end{sloppypar}

In order to examine the impact of the \(K\)-factor approximation on state-of-the-art PDF analyses,
we have utilised the open-source NNPDF fitting framework~\cite{NNPDF:2021uiq} to perform a series of fits under different assumptions:
DY theory predictions are varied between the exact \NNLO grids and the \(K\)-factor approximation, and the data entering the fits either include the full NNPDF4.0 dataset or a reduced dataset based solely on collider DY measurements (``DY only fit'').
The latter fit based on the restricted data is intended to act as the worst-case scenario where \emph{all} datasets are impacted by the approximation in the theory predictions.
To closely mimic the procedure of an actual PDF fit, we have computed the \(K\)-factor with the closest PDF, in this case NNPDF4.0~\cite{NNPDF:2021njg}.

Note that for the results presented in this section, aside from the grids, all other conditions and settings are kept identical between the two fits, including initialisation and seeding of random numbers.

\begin{figure}
  \centering
  \includegraphics[width=0.95\linewidth]{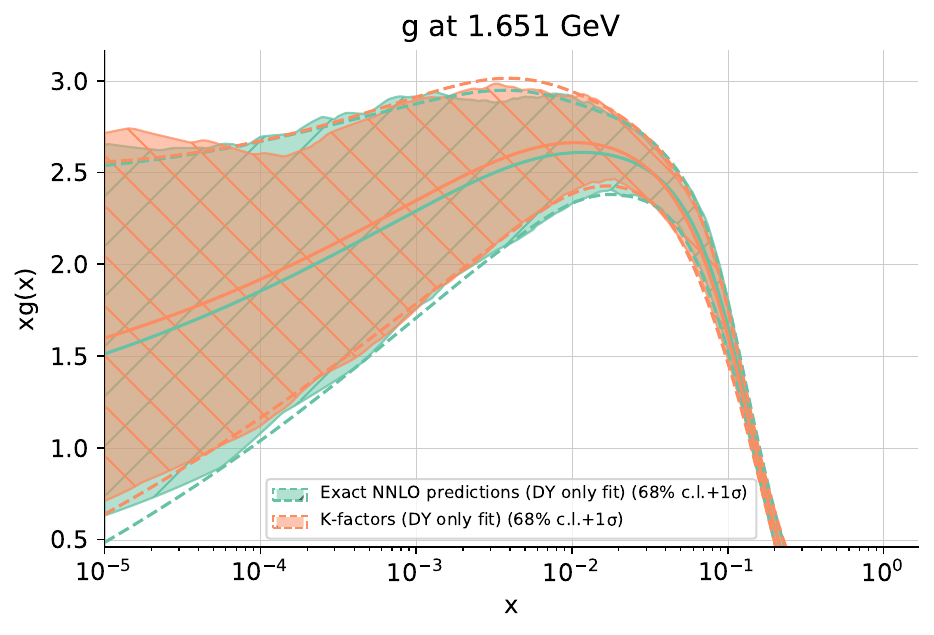}
  \includegraphics[width=0.95\linewidth]{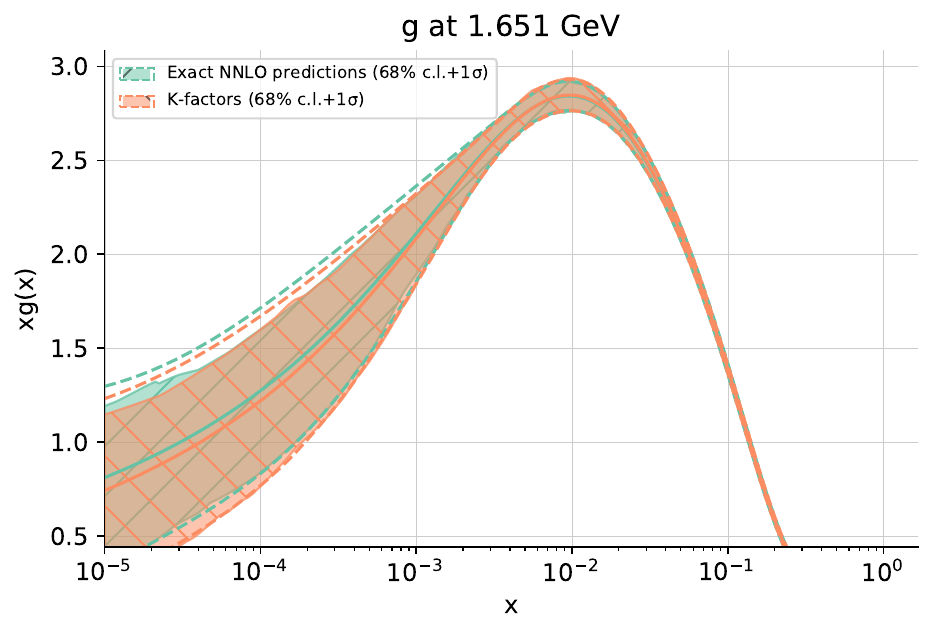}
  \caption{
    Comparison of the gluon PDF between two PDF fits, one with exact NNLO grids and the other based on the \(K\)-factor approximation.
    The fit is repeated using only DY data~(top) and the global dataset~(bottom).
  }
  \label{fig:gluononly}
\end{figure}

\begin{figure}
  \centering
  \includegraphics[width=0.95\linewidth]{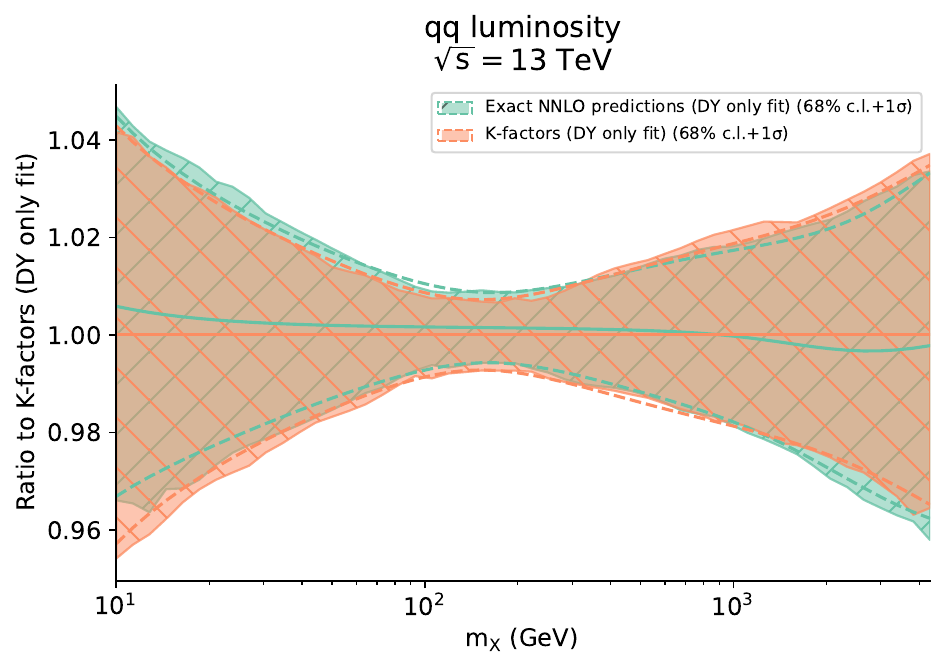}
  \includegraphics[width=0.95\linewidth]{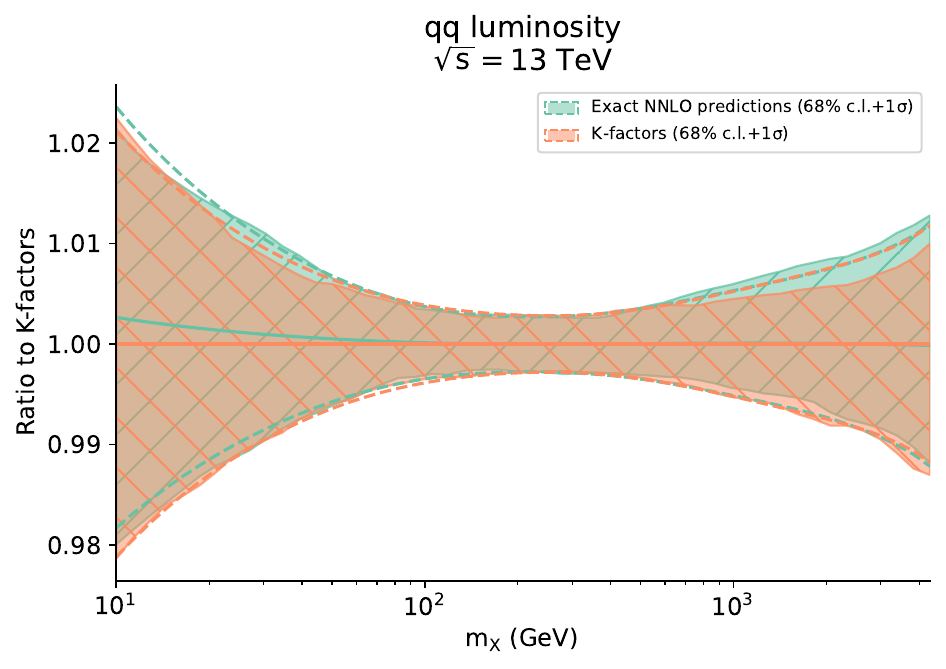}
  \caption{
    Comparison of the $qq$ luminosity between two PDF fits, one with exact NNLO grids and the other based on the \(K\)-factor approximation.
    The fit is repeated using only DY data~(top) and the global dataset~(bottom).
  }
  \label{fig:fitlumi}
\end{figure}

In Fig.~\ref{fig:gluononly} we compare the effect of the \(K\)-factor approximation in the context of a global fit by explicitly showing the gluon distribution at the fitting scale.
We consider two situations, a fit based exclusively on DY data from LHC and Tevatron~(top) and a fit using a global dataset~(bottom).
In the first case the impact of the \(K\)-factor approximation is visible even in the data region, with a small shift in the data region which however is below a percent and safely well within the uncertainties of the determination.
In the global fit instead we consider the NNPDF4.0 settings, i.e., all DIS data is exact up to NNLO and double-hadronic data instead uses the \(K\)-factor approximation.
The better constraints introduced by the global dataset make it so that when using exact grids for the DY data, the effect is completely negligible and only visible in the extrapolation region, where instabilities introduced by the small numerical differences between the \(K\)-factor and exact approximation won't be compensated by the fit.
While we only consider the gluon in Fig.~\ref{fig:gluononly}, the same is true for other partons.

In Fig.~\ref{fig:fitlumi} we display a comparison of the \(qq\) luminosity in the same scenarios as Fig.~\ref{fig:gluononly}.
We observe how the impact of the \(K\)-factor approximation affects the entire range of the phase space in a DY only fit, while the inclusion of the rest of the datasets commonly considered in a global fit eliminates the impact (note the different range on the $y$-axis).

From the comparison of the different fits presented in this section, we can conclude that the \(K\)-factor approximation is a valid and safe approximation for the DY process in the context of PDF fits.
Even in the specially tuned scenario where the impact of the approximation was maximised by restricting the fit data only to hadron-collider Drell--Yan measurements, the impact on the resulting PDFs was found to be minimal and well within uncertainties.

\section{Conclusions and outlook}
\label{sec:conclusions}

In this note we have reported on a new interface between the \pineappl interpolation grid library and the \NNLO parton-level Monte Carlo generator \nnlojet.
This interface was used to produce interpolation grids for a wide range of DY measurements performed at Tevatron and the LHC that commonly enter global PDF analyses.

These grids were used to gain further insights into the accidental cancellations that occur in the Drell--Yan process at \NNLO, where strong correlations induced by the DGLAP evolution within the singlet sector were identified as a main driver of the cancellations.
We further performed a detailed study of the \(K\)-factor approximation that was employed in the PDF fits so far.
While few-\% variations are found from the variation of PDF sets in the \NNLO cross section, the \(K\)-factor is found to be very stable with only changes at the per-mille level.
A set of PDF fits based on varying theory predictions as well as fit data further support the conclusion that the impact of the \(K\)-factor approximation is minimal and well below the quoted PDF uncertainties.

The independence of PDF fits on the \(K\)-factor approximation for the DY data is a non-trivial consequence of the observations detailed in sections~\ref{sec:cancellation} and~\ref{sub:K-fac}.
While there are big channel-by-channel cancellations, these are dominated by DGLAP evolution.
While PDF determinations might differ widely in data and methodology, at a fixed order the evolution is fixed, and so the pattern of cancellations is preserved regardless of the PDF in use.

The grids provided in this work not only allowed to establish the validity of the \(K\)-factor approximation for DY at \NNLO but also paves the way towards incorporating \N3\LO predictions into PDF fits in the future.
Full \N3\LO grids are still beyond reach due to computation costs, however, the \NNLO grids provide one ingredient to construct approximate \N3\LO predictions based on \N3\LO \(K\)-factors.

\section*{Acknowledgements}
C.S.\ is supported by the German Research Foundation (DFG) under reference number DE 623/6-2.

\bibliographystyle{utphys}
\bibliography{dy_grids}

\end{document}